# A NEW APPROACH TO QUANTUM GRAVITY
*AN OVERVIEW*

Sarah B.M. Bell, John P. Cullerne, Bernard M. Diaz
*Department of Computer Science I.Q. Group, The University of Liverpool, Liverpool, L69 7ZF*

Abstract:    We quantise General Relativity for a class of energy-momentum-stress tensors.

## 1. The new method of quantisation

### 1.1 THE GENESIS OF A CURVED SPACETIME

It has been generally believed that QED does not allow a classical spin for the particle described. However, it has now been shown that QED permits the spin of the particle to behave like a four-vector (Bell et al. 2000a&b). This was done by mapping the Dirac and photon equation into a new form, the versatile form,

$$( \underline{D} - i e \underline{\tilde{A}} ) \underline{\Phi} = \underline{\Phi}\, \underline{M}, \tag{1}$$

where e is the charge on the electron. This permits the application of QED as a quantum version of General Relativity.

We will study one particular sort of curved spacetime here which we will use as the building block for all the others. In this section we discuss its extrinsic description and relationship to the versatile Dirac and photon equations. We define the curvature in terms of co-ordinate systems. We shall call the original flat spacetime that applies to equation (1) the spacetime of *the Large observer*. We may co-ordinate the $x_1$, $x_2$ and $x_0$, $x_3$ planes of the Large observer with polar co-ordinates, $r_1$, $\theta_1$, and $r_2$, $\theta_2$, where we have $x_1 = r_1 \sin(\theta_1)$, $x_2 = r_1 \cos(\theta_1)$, $x_0 = r_0 \sin(\theta_0)$, $x_3 = r_0 \cos(\theta_0)$ and

$$s_1' = r_1 \theta_1, \quad s_0' = r_0 \theta_0, \tag{2}$$

and $s_1'$, $s_0'$ are the arcs corresponding to $r_1$, $r_0$ and $\theta_1$, $\theta_0$. We introduce two further observers who we shall call *the Medium observer* and *the Small observer*. We give the first the co-ordinate system ( $x_0, s_1, r_1, x_3$ ), while the second is given the co-ordinate system ( $s_0, s_1, r_1, r_0$ ). The spacetime of these observers is curved in such a way that equation (2) is replaced by

$$s_1 = R_1 \theta_1, \quad s_0 = R_0 \theta_0, \tag{3}$$

where $R_1, R_0$ are constant. $s_1, s_0$ are arcs as before. Both of equations (3) apply for the Small observer, while the first of equations (3) and the last of equations (2) apply to the Medium observer. The spacetime of the Medium observer is associated with the circular orbits of the Bohr atom while that of the Small observer is associated with the elliptical orbits of the Sommerfeld atom.

The Dirac equation (1) and the photon equations are transformed into the curved spacetimes of the Medium and Small observers using variables, ( $x_0, s_1, r_1, x_3$ ) for the Medium observer and ( $s_0, s_1, r_1, r_0$ ) for the Small observer. The quaternion matrices are transformed to lie along the arc and radius. They are then also transformed to the curved spacetime of the Medium and Small observers so that they appear constant for every location in their spacetimes. The result for the Medium observer is
$\underline{D}_1 + \underline{D}_2 = \underline{i}_{s1} \partial / \partial s_1 + \underline{i}_{r1} \partial / \partial r_1$, $\underline{D}_0 + \underline{D}_3 = i \underline{i}_0 \partial / \partial x_0 + \underline{i}_3 \partial / \partial x_3$, with $\underline{D}_0$ and $\underline{D}_3$ unchanged, while for the Small observer $\underline{D}_1 + \underline{D}_2 = \underline{i}_{s1} \partial / \partial s_1 + \underline{i}_{r1} \partial / \partial r_1$, $\underline{D}_0 + \underline{D}_3 = i \underline{i}_{s0} \partial / \partial s_0 + \underline{i}_{r0} \partial / \partial r_0$. We have ( $\underline{i}_{s0}, \underline{i}_{r0}$ ) = ( $\underline{i}_0, \underline{i}_3$ ), ( $\underline{i}_{s1}, \underline{i}_{r1}$ ) = ( $\underline{i}_1, \underline{i}_2$ ), and for the reflector matrices $\underline{i}_\mu = \underline{i}_\mu ( \underline{i}_\mu, \underline{i}_\mu^\ddagger )$. Similar expressions apply to $\underline{M}$ and $\underline{A}$. However, while the spacetime of the Medium and Small observers is locally flat, it is globally curved and this will affect any variable that depends on volume. The potential depends on a probability current per unit volume and therefore $\underline{A}$ does not refer to the potential the Medium or Small observer will see. Equations (2) and (3) are used to find the volume seen by the Medium or Small observer and the photon equation to find the corresponding potential. This calculation shows that an inverse distance potential seen by the Large observer becomes a constant potential in the spacetimes of both the Medium and Small observers.

More details are provided by Bell et al. (2000f).

1.2 THE BOHR-SOMMERFELD ATOM

We may now solve the Dirac equation amended as described in section 1.1 for the spacetime of the Medium observer to show the derivation of the Bohr atom as a consequence of QED. Our version of Bohr's equations, unlike his original treatment, will be relativistically correct. We suppose that the Bohr electron making up the one-electron atom has a velocity in the $s_1$ direction, that is, the electron has a circular orbit round the proton. We use the rest frame of the proton. We discover that

$$\left. \begin{array}{r} \phi_1 = \exp \{ i ( \tilde{v} \tilde{x}_0 + \mu s_1 ) \} \\ \phi_2 = \{ ( i \tilde{v} - i \underline{i}_{s1} \mu - i e \tilde{A} ) M^{-1} \} \times \exp \{ i ( \tilde{v} \tilde{x}_0 + \mu s_1 ) \} \\ \tilde{m}^2 = ( \tilde{v} - e \tilde{A} )^2 + \mu^2, \end{array} \right\} \tag{4}$$

where we have set $h / 2\pi$ to 1, $\tilde{m} = ( m_e / i )$ with $m_e$ the mass of the electron, $\tilde{x}_0 = x_0 / i$, and we have a plane wave solution $\underline{\Phi}( \phi_1, \phi_2 )$, where $i \tilde{v}$ is the frequency and $\mu$ the wave number. $\tilde{A} / i$ is the potential due to the proton in the Medium observer's spacetime. We have

$$\tilde{A} = i\,e\,/\,R_1, \tag{5}$$

where e is the charge on the proton. $R_1$ is the Bohr radius at which both the Large and Medium observers see the same potential. $R_1$ is a constant. We see that the interaction between the proton and electron has introduced no more than a change in the vacuum potential and swap where co-ordinates ( $x_1$, $x_2$ ) are exchanged for co-ordinates ( $s_1$, $r_1$ ).

We may derive the Bohr equations from the solution to the Dirac equation, equation (4). We see from the third of equations (4) that there is an equivalent free electron with wave number μ and frequency

$$\tilde{\eta} = \tilde{\nu} - e\,\tilde{A}. \tag{6}$$

This is the Bohr electron. We require that the wave function to be single-valued from the point of view of the Large observer. De Broglie's relation between μ and the momentum of the Bohr electron then provides

$$\tilde{m}\,\tilde{v}\,R_1\,/\,\sqrt{(1 + \tilde{v}^2)} = n_\theta\,(h\,/\,2\,\pi), \tag{7}$$

where ( $-i\,\tilde{v}$ ) is the velocity of the electron, we have restored $h/2\pi$ and $n_\theta$ is an integer. This is the first of Bohr's equations. Equations (5) and (6), and de Broglie's relation between η and the energy of the Bohr electron provides

$$\tilde{\nu} = \tilde{m}\,/\,\sqrt{(1 + \tilde{v}^2)} + i\,e^2\,/\,R_1 \tag{8}$$

However, there is another way of obtaining the total energy and that is to consider the geometry of the path of the Bohr electron. From the Bohr electron's point of view the mass is $i\,\tilde{m}$ and the electron is in a potential field. However, from the point of view of the atom as whole there is no potential field and $i\,\tilde{m}$ is the frequency associated with the mass of the atom, $i\,\tilde{v}$, travelling at velocity $\tilde{v}$ with respect to the atom's rest frame. From de Broglie's relation for the atom, we must therefore have $\tilde{m} = \tilde{v}^{b} / \sqrt{(1 + \tilde{v}^{b-2})}$. This leads to

$$\tilde{\eta}\,\delta\tilde{x}_0 + \mu\,\delta s_1 = \tilde{\nu}\,\delta\tilde{x}_0, \tag{9}$$

and to

$$i\,e^2\,/\,R_1 = \tilde{m}\,\tilde{v}^2\,/\,\sqrt{(1 + \tilde{v}^2)}, \tag{10}$$

which is Bohr's second equation.

The total angular momentum of the Bohr atom is made up of an integer contribution from the orbital angular momentum, the half integer spin of the electron and a half integer contribution from the Berry phase (Anandan 1992) the electron gains each orbit because the orbit can be considered as a closed path containing the nucleus.

We next derive the Sommerfeld model as a consequence of QED. This time we want to replace co-ordinates ( $x_0$, $x_3$ ) with ( $s_0$, $r_0$ ) and take on the point of view of the Small observer. We adopt $s_1$ as our temporal co-ordinate with ( $x_0$, $x_3$ ) spatial so that the situation becomes exactly analogous to the our derivation of the Bohr atom above. We can do this because the Dirac equation (1) can be shown to have a suitable tachyon form. We suppose a tachyon particle and anti-particle emerge, if the atom is suitably excited, in the form a bound state. We call the new particles the *Green electron* and *Green positron*. We shall call the bound state they form *the Green atom*. The emergence of the Green atom causes a transition from the spacetime of the Medium observer to that of the Small observer. The wavefunctions of both the Bohr electron and Green electron must be single-valued along the new loop $s_0$ and we may deduce two similar equations to (7). This is sufficient to enable us to calculate the tachyon equivalent of the rest mass of the Green electron and positron, $i\,m^{g\sim} = -(m^\sim / v^{g\sim})(n_r / n_\theta)$, where $n_r$ is an integer called the azimuthal quantum number. We may also reproduce Bohr's equations for the Green atom assuming $s_1$ is the temporal co-ordinate and the Green positron stationary. The velocity of the Green electron as it appears in Bohr's equations is $i\,v^{g\sim}$. We next swap $s_1$ for $s_0$ as the temporal axis by swapping energy and momentum eigenvalues for the Green electron and positron and discover the total contribution the Green atom makes to the frequency of the complete system,

$$E_G = -m^{g\sim} v^{g\sim}. \tag{11}$$

We add the contributions of the Bohr electron and Green atom. This gives us the total frequency and momentum of the whole stucture in the spacetime of the Small observer. Repeated use of analogues to equation (9) allow us to transform these to the spacetime and frame of interest, that of the Large observer and ($x_0, x_1, x_2, x_3$), where we retrieve the Sommerfeld expression for the energy levels of the atom.

More details are provided by Bell et al. (2000e&f).

1.3 THE NEW QUANTIZATION

In section 1.2 we sketched the derivation of the Bohr and Sommerfeld models of the atom from QED. Here we do the inverse, derive QED from the Bohr and Sommerfeld models. We do not need to do more than derive QED from the Bohr model because the extra Sommerfeld energy levels appear when the Bohr model is applied a second time, using the tachyonic Green atom. It is well known that the inverse square of distance law for the strength of the electrical field gives rise to the photon equation in differential form. We extend this using a similar method to the interaction of another particle with the potential field, A, described by the photon equation.

We assume the Bohr equations, (7) and (10), apply to an electron interacting with an spherical volume of constant charge density responsible for the potential field, A, at point b in the spacetime of the Medium observer. The Bohr electron orbits round the surface of the spherical volume at the Bohr radius and we may arrange for the electron to make as many orbits as may be required to consider the electron bound whatever the total interaction may describe. We do this by letting the spherical volume decrease in size. As the volume becomes smaller the mass of the electron, $m_e$, will increase along with the charge, e, as in renormalisation.

If we add equation (8) describing energy conservation to the Bohr equations we may assemble a wave function that describes the Bohr electron in a manner similar to equation (4). As the spherical volume becomes infinitesimal this wavefunction will satisfy the Dirac equation at point b from the point of view of the Medium observer. However, the steps we took to get from the Dirac equation in the spacetime of the Large observer to the Dirac equation in the spacetime of the Medium observer are reversible. We may return to a Dirac equation and solution with the same frequency term, $\tilde{\nu}$, and equivalent angular momentum at point b in the spacetime of the Large observer.

If Bohr's equations are valid for infinitesimal volumes at all points in the field then the Dirac equation must hold everywhere too and we therefore find that the full quantum theory is QED. We note that we have not mentioned the global boundary conditions or constrained the magnitude of the wavefunction in our discussion. The magnitude of the wave function would vary as required by the global boundary conditions.

We find that if we negate both charges on our interacting bodies and replace the potential, A by ( A – 1 ) QED continues to apply.

More details are provided by Bell et al. (2000e).

## 2. Application of the new method to gravity

### 2.1 BOHR'S FIRST EQUATION

We now derive Bohr's two equations for quantum gravity. With electromagnetism we are in the position of the Large observer and the spatial loop, $s_1$, provides the initial bound state, with the temporal loop, $s_0$, giving extra fine structure. For gravity we are in the position of the Medium observer, and the first loop we see is temporal. Since we want to distinguish the gravitational hydrogen atom from the electromagnetic equivalent we will call the former *the Thalesium atom.* We will call the equivalent of the Green electron for a Thalesium atom *a Geotron.*. We will identify the point of view of the Green positron with that of the Thalesium atom. As we did above, we will assume the point of view of the Thalesium atom initially and assume our temporal co-ordinate is spatial. We therefore have the frame and spacetime of the Thalesium atom as being equivalent to that of the Medium observer except that a spatial and temporal co-ordinate are swapped, and the frame and spacetime of the Geotron as being equivalent to that of the Small observer except that a spatial and temporal co-ordinate are swapped and that the Geotron is a moving frame relative to the Thalesium atom.

Spacetime appears flat to the Medium observer with his co-ordinates ( $x_0$, $s_1$, $r_1$, $x_3$ ) Cartesian since all the curvature produces is a variation in the vacuum potential. We will change our nomenclature and take the metric to be,

$$dt^2 = dx_0^2 - dx_1^2 + dx_2^2 + dx_3^2, \qquad (12)$$

where we have assumed the point of view of the Thalesium atom and are treating $x_1$ as temporal. We shall sometimes need the variant

$$dt^2 = r^2 d\theta^2 - dx_1^2 + dx_2^2 + dr^2, \qquad (13)$$

where ( r, θ ) are polar co-ordinates.

The metric we shall use to describe the curved spacetime of the Geotron is

$$d\tau^2 = (a/r) dt^2 + (r/a) dr^2 - dx_1^2 + dx_2^2, \qquad (14)$$

where we have set the gravitational constant to unity, t is the temporal co-ordinate and the analogue of $s_0$ in section 1. However, we are treating $x_1$ as temporal. a is twice the mass of the source responsible for the gravitational attraction, which takes on the role the Green positron played in the electromagnetic case. We suppose that equation (14) is an intrinsic description of the curvature we now link to QED. We want to consider a temporal loop where the Geotron is bound to the Thalesium atom and in a circular orbit. We therefore use metric (13) and set $dt = r\, d\theta$, where $d\theta$ is the change in angle according to the Thalesium atom. Equation (14) becomes

$$d\tau^2 = a\, r\, d\theta^2 + (r/a)\, dr^2 - dx_1^2 + dx_2^2. \qquad (15)$$

We consider a circuit of the Geotron round the centre of attraction in the plane $r, \theta$ in the metric given by equation (15). From the point of view of the Geotron position four-vector of the Geotron does not change direction, but the co-ordinate frame of the Thalesium atom does. So we can consider the position four-vector parallel transported as for the Berry phase. Let the two-vector giving the position of the Geotron be $V = (V^r, V^\theta)^T$. Following Martin (1995) and assuming the Geotron orbits at a fixed radius r, we then find

$$\theta_S = (1/2)\sqrt{(a/r)}\,(\tau/r), \quad V^r = \sin\theta_S, \quad V^\theta = (-1/a)\cos\theta_S. \qquad (16)$$

$\theta_S$ is also the angle the position vector of the Geotron makes relative to the Thalesium atom. We require a steady state in which the system returns to start position after a cycle. In that case after n circuits, when from the point of view of the Thalesium atom $\tau = 4\pi\, r\, n$, $\theta_S$ must be an integer multiple of $2\pi$, $\theta_S = 2\pi\, m$, with both n and m integer. Although the spin is integer from the point of view of the Thalesium atom because the Berry phase is added, from the point of view of the Geotron the spin is half integer as is that of the electron which is why we have added an extra factor of 2 for $\tau$. From the Thalesium atom's viewpoint the path length corresponds to an angle $2\,\theta_M = 4\pi\, n$. We have

$$d\theta_S / d\theta_M = m/n. \qquad (17)$$

Substituting for $\tau$ and $\theta_S$ in equations (16) we see this imposes a condition on r, $r = R$, where R is a constant and

$$\sqrt{(a/R)} = m/n. \qquad (18)$$

R is the equivalent of the Bohr radius and we shall call it that, defining it exactly in section 2.2. We would like to know the interval as the Thalesium atom sees it. Using metric (13) we obtain for the interval

$$dt_M = r\, d\theta_M. \qquad (19)$$

We would also like to know the interval, $dt_S$, as the Geotron sees it. Substituting from equation (19) into metric (14) and using equation (17) and (18) we obtain

$$dt_S = d\tau = r\, d\theta_S. \qquad (20)$$

We see that the Geotron is also entitled to see his spacetime as flat. Equation (18) provides information which will allow us to derive Bohr's first equation.

## 2.2 BOHR'S SECOND EQUATION

The Lagrangian associated with equation (14) gives us the acceleration of the Geotron for the orbit of the Geotron we discussed in section 2.1. The acceleration is

$$d^2r / d\tau^2 = - a / ( 2 r^2 ). \qquad (21)$$

Next we calculate the acceleration seen by an observer with the frame of the Thalesium atom in his flat spacetime. We apply General Relativity to the problem (Kenyon 1990) and consider the covariant derivative for metric (12) and the equation for the geodesic in flat spacetime. From these we obtain

$$d^2r / dt^2 = - v_g^2 / r, \quad v_g = r ( d\theta_M / dt_M ) = r ( d\theta_S / dt_S ), \qquad (22)$$

where $v_g$ is equivalent to the velocity of the Geotron and we have used equations (19) and (20) for the last two equations of (22). We see from equations (22) that it does not matter whether we are in frame and spacetime of the Thalesium atom or in the frame and spacetime of the Geotron. We obtain from equations (22) and (21)

$$v_g^2 = a / ( 2 R ). \qquad (23)$$

From equations (18) and (23) we may formulate the Bohr equations in exactly the same form as they took for QED as described by Bell et al. (2000c&d).

## 2.3 QUANTUM GRAVITATIONAL DYNAMICS

We have been taking the point of view of the Thalesium atom or Geotron for our calculations above. We now wish to assume the point of view of the Small observer for whom the circular loop is temporal. We calculated the energy contribution for such a temporal loop in section 1.2, equation (11). For the current purpose we re-state the equation as $E_n = m_g v_g$. We may also calculate the gravitational energy in the standard way (Martin 1995) from the metric (14) and Lagrangian. The two methods agree on the form of the result. On taking up the position where the loop is temporal our metric for the Medium and Small observer become
$dt^2 = - dx_0^2 + dx_1^2 + dx_2^2 + dx_3^2$, and $d\tau^2 = - ( a / r ) dt^2 + ( r / a ) dr^2 + dx_1^2 + dx_2^2$.

We consider the state of the Thalesium atom. For QED there is no preferred plane for the angular momentum when we are not in an eigenstate. All possible orbits exist as a superposition of states and the atom appears spherically symmetrical. We assume the same for the Thalesium atom. The metric of the Small observer becomes $d\tau^2 = - ( a / r ) dt^2 + ( r / a ) dr^2 + r^2 d\zeta^2 + r^2 \sin^2( \zeta ) d\xi^2$, where r labels a sphere of area $4\pi r^2$ and $\zeta$ and $\xi$ describe colatitude and longitude. The Bohr equations can be derived in an identical manner from this metric as well, provided that the first two terms are set positive.

We apply the method of quantisation we used in section 1.3. We find that QED is the full quantum theory. Moreover, we can find the equivalent of the electromagnetic potential from the Bohr equations. In the frame of the Geotron it is $A = a / ( 2 r )$. Since we are discussing gravity, while keeping the equation itself the same, we will rename the photon equation the graviton equation. If we take the global point of view the potential will be a full four-vector. For the metric, we must set

$$g_{tt} = 2 A(x_\mu), \quad g_{rr} = \{2 A(x_\mu)\}^{-1}, \quad g_{\zeta\zeta} = r(x_\mu)^2, \quad g_{\xi\xi} = r(x_\mu)^2 \sin^2(\zeta), \quad (24)$$

for the local infinitesimal volume and then Lorentz transform into the global frame by some $Z(x_\mu)$.

Equation (24) covers Schwarzschild metric for the spacetime applying to a spherically symmetric source of gravitation (Martin 1995). We may therefore move to an interpretation of our general metric (24) from the point of view of General Relativity. We impose a change the potential of + 1 for each infinitesimal spherical volume in the local rest frame and negate its charge. We negate the charge of the body interacting with the field. QED continues to apply, but each spherical volume has, in the local rest frame, the Schwarzschild metric. This means that, if we have the redefintion of metric (24) applying, at each location and time we have an infinitesimal system composed of two bodies interacting according to the theory of General Relativity. The quantum condition discussed in section 1.3 is still in place. At this point it can be shown that there is a one-to-one correspondence between the current density of QED, $J_\mu$, and the energy-momentum-stress tensor, $T_{\mu\nu}$, of the Einstein equation for the permissible metrics of type (24) plus local Lorentz transformation $Z(x_\mu)$.

**Acknowledgement**
One of us (Bell) would like to acknowledge the assistance of E.A.E. Bell.